\documentclass[10pt, final, journal, twoside]{IEEEtran}

\usepackage{cite}
\usepackage{mathtools}
\usepackage{epsf}
\usepackage{commath}
\usepackage{amsmath}
\usepackage{amssymb}
\usepackage{amsfonts}
\usepackage{graphicx}
\usepackage[percent]{overpic}

\newcommand{\dmax}{d_{\mathrm{max}}}
\newcommand{\rmax}{r_{\mathrm{max}}}
\newcommand{\dtot}{d_{\mathrm{tot}}}

\usepackage{scalerel}
\usepackage{tikz}
\usetikzlibrary{svg.path}
\definecolor{orcidlogocol}{HTML}{A6CE39}
\tikzset{
  orcidlogo/.pic={
    \fill[orcidlogocol] svg{M256,128c0,70.7-57.3,128-128,128C57.3,256,0,198.7,0,128C0,57.3,57.3,0,128,0C198.7,0,256,57.3,256,128z};
    \fill[white] svg{M86.3,186.2H70.9V79.1h15.4v48.4V186.2z}
                 svg{M108.9,79.1h41.6c39.6,0,57,28.3,57,53.6c0,27.5-21.5,53.6-56.8,53.6h-41.8V79.1z M124.3,172.4h24.5c34.9,0,42.9-26.5,42.9-39.7c0-21.5-13.7-39.7-43.7-39.7h-23.7V172.4z}
                 svg{M88.7,56.8c0,5.5-4.5,10.1-10.1,10.1c-5.6,0-10.1-4.6-10.1-10.1c0-5.6,4.5-10.1,10.1-10.1C84.2,46.7,88.7,51.3,88.7,56.8z};
  }
}
\newcommand\orcidicon[1]{\href{https://orcid.org/#1}{\mbox{\scalerel*{
\begin{tikzpicture}[yscale=-1,transform shape]
\pic{orcidlogo};
\end{tikzpicture}
}{|}}}}
\usepackage{hyperref} 

\begin{document}

\title{{\LARGE Improved Coefficients for the Karagiannidis--Lioumpas\\Approximations and Bounds to the Gaussian $Q$-Function}}
\author{Islam~M.~Tanash \orcidicon{0000-0002-9824-6951} and Taneli~Riihonen \orcidicon{0000-0001-5416-5263},~\IEEEmembership{Member,~IEEE}%
\thanks{Manuscript received November 19, 2020; revised December 18, 2020 and January 8, 2021; accepted January 9, 2021. Date of publication January 18, 2021; date of current version \today. This work was partially supported by the Academy of Finland under Grant 326448. The associate editor coordinating the review of this letter and approving it for publication was A.-A.~A.~Boulogeorgos. ({\em Corresponding author: Islam~M.~Tanash.})}%
\thanks{The authors are with Tampere University, Tampere 33720, Finland (e-mail: islam.tanash@tuni.fi; taneli.riihonen@tuni.fi).}%
\thanks{Digital Object Identiﬁer 10.1109/LCOMM.2021.3052257}}

\markboth{IEEE COMMUNICATIONS LETTERS, in press}{Tanash and Riihonen: Improved Coefficients for the Karagiannidis--Lioumpas Approximations and Bounds to the $Q$-Function}

\maketitle

\begin{abstract}
We revisit the Karagiannidis--Lioumpas (KL) approximation of the $Q$-function by optimizing its coefficients in terms of absolute error, relative error and total error. For minimizing the maximum absolute/relative error, we describe the targeted uniform error functions by sets of nonlinear equations so that the optimized coefficients are the solutions thereof. The total error is minimized with numerical search. We also introduce an extra coefficient in the KL approximation to achieve significantly tighter absolute and total error at the expense of unbounded relative error. Furthermore, we extend the KL expression to lower and upper bounds with optimized coefficients that minimize the error measures in the same way as for the approximations.
\end{abstract}


\section{Introduction}
\label{sec:Introduction}

\IEEEPARstart{K}{aragiannidis and Lioumpas} presented in~\cite{Karagiannidis-Lioumpas-2007Aug} a relatively tight, yet analytically tractable, approximation for {\em the} Gaussian $Q$-function\cite{Aggarwal-2019Q3} as follows:
\begin{align}
\begin{split}
Q(x)
&\triangleq \frac{1}{\sqrt{2\pi}} \int_{x}^{\infty} \exp\left(-{\textstyle \frac{1}{2}} t^2\right)\,\mathrm{d}t\\
&\approx a\exp\left(-b\,x^2\right)\cdot\frac{1-\exp\left(-c\,x\right)}{x}
\triangleq \tilde{Q}(x)
\label{eq:Q_and_Qtilde}
\end{split}
\end{align}
for which their original study sets $(a,b,c) = (\frac{1}{B\sqrt{2\pi}}, \frac{1}{2}, \frac{A}{\sqrt{2}})$ and proposes for error minimization example coefficient values $A = 1.98$ and $B = 1.135$ rendering $(a,c) \approx (0.3515, 1.4001)$.

Despite drawing some criticism~\cite{Dyer-Dyer-2008Apr} shortly after publication, {\em the} `Karagiannidis--Lioumpas (KL) approximation' has gradually established itself as one of the most usable substitutes for the Gaussian $Q$-function in communication theory problems and the paper~\cite{Karagiannidis-Lioumpas-2007Aug} has received a large number of citations; it is only fitting to begin calling the expression after its inventors.

A diverse set of applications for the KL approximation can be found in~\cite{Potter-Venayagamoorthy-Kosbar-2010Feb,Tan-Le-2011Oct,Wu-Mehta-Molisch-Zhang-2011Dec,7544526,8626200} to name but a few prominent articles. In general, the approximation is often used in the calculation of average bit or symbol error probability as a tractable replacement for the Gaussian $Q$-function such that analysis can be carried out and completed in a closed form at the cost of making results tight approximations instead of exact ones. This usually involves integrating something like, e.g., $f(\tilde{Q}(x(\gamma)))$, where even simple functions $f(q)$ and $x(\gamma)$, which are derived from the communication system under study, may forbid exact analysis using the actual $Q$-function\cite{Chiani-Dardari-Simon-2003Jul,LopezBenitez-Casadevall-2011Apr,Tanash-TCOM1}. One should note especially that the approximation is always used in an intermediate step of analytical derivations and it is not meant for numerical probability computations per se --- instead, rational Chebyshev functions~\cite{Cody-1969Jul} 
are perfect to that end.

This Letter is inspired by the fact that the original study~\cite{Karagiannidis-Lioumpas-2007Aug} presents explicit values of $a$ and $c$ for {\em only one approximation} (which has low integrated total error when $b = \frac{1}{2}$, to be exact). However, the KL approximation family is actually much more versatile, whereby new coefficients can be acquired in terms of other criteria for better accuracy depending on the application. The KL expression can be also repurposed to achieve lower and upper bounds (that are also tight approximations) and, in certain cases, coefficients admit explicit values. Furthermore, by introducing the extra coefficient $b$ in (\ref{eq:Q_and_Qtilde}) that originally was $b=\frac{1}{2}$ and permitting $b < \frac{1}{2}$, we achieve significantly improved accuracy in terms of absolute and total error.

The objective of this Letter is to  apply the KL expression of the $Q$-function to derive improved approximations and bounds which are global and tight over $x\geq0$ by optimizing the coefficients $(a,b,c)$ in respect to their minimum global absolute or relative error or minimum integrated total error. Like\cite{Tanash-TCOM1} for another popular expression\cite{Chiani-Dardari-Simon-2003Jul}, we present new formulation that minimizes the maximum global error of (\ref{eq:Q_and_Qtilde}) by constructing a set of equations, which describes the corresponding error function, and solve them numerically to find the optimized coefficients. The total error is optimized with exhaustive search for reference. In general, when optimizing one of the three criteria, better performance will be achieved at the expense of decreased accuracy in terms of the others.

The new coefficients solved herein are applicable as one-to-one replacements for the original ones of \cite{Karagiannidis-Lioumpas-2007Aug} adopted into the analysis of~\cite{Potter-Venayagamoorthy-Kosbar-2010Feb,Tan-Le-2011Oct,Wu-Mehta-Molisch-Zhang-2011Dec,7544526,8626200} and many other studies. Literature is rich in approximations/bounds for the $Q$-function and, typically, the application's mathematics define, which ones are tractable for it. Whenever (\ref{eq:Q_and_Qtilde}) is preferred, our coefficients offer variety to tailor accuracy for the application or to use bounds. The tractable series expansion of the KL expression proposed in~\cite{Isukapalli-Rao-2008Sep} can likewise be used with these substitutes (but without guarantee that bounds remain global). Consequently, they are useful also in the contexts of, e.g., \cite{Isukapalli-Rao-2010Apr,Zhou-Li-Lau-Vucetic-2010Nov,Seyfi-Muhaidat-Liang-2011Jan,Seyfi-Muhaidat-Liang-Dianati-2011Dec,Banani-Vaughan-2012Oct,Choi-Hanzo-2013Jul,7880559} and many others.

\section{Preliminaries}
\label{sec:Preliminaries}

The case $x \geq 0$ is presumed throughout this Letter with little loss of generality because the relation $Q(x) = 1 - Q(-x)$ extends all the considered functions to the negative real axis. In fact, this is the main motive for optimizing approximations and bounds also subject to an additional constraint $\tilde{Q}(0)=\frac{1}{2}$ that makes their extensions continuous at the origin like $Q(x)$.

This study solves optimized approximations and bounds for three criteria and for combinations thereof, viz.\ $\min_{(a,b,c)} \dmax$ (`minimax absolute error'), $\min_{(a,b,c)} \rmax$ (`minimax relative error') and $\min_{(a,b,c)} \dtot$ (`integrated total error'~\cite{Karagiannidis-Lioumpas-2007Aug}), where\\
\mbox{$\displaystyle \dmax \triangleq \max_{x \geq 0} |d(x)|,\,
\rmax \triangleq \max_{x \geq 0} |r(x)|,\,
\dtot \triangleq \int_{0}^{\infty} \hspace{-5pt}|d(x)|\,\mathrm{d}x,$}\\
and the error functions are defined as
\begin{align}
\label{eq:abs_error}
d(x) &\triangleq \tilde{Q}(x) - Q(x),\\
\label{eq:rel_error}
r(x) &\triangleq \frac{d(x)}{Q(x)} = \frac{\tilde{Q}(x)}{Q(x)} - 1.
\end{align}
For baseline reference, the coefficients originally given in~\cite{Karagiannidis-Lioumpas-2007Aug} render $\dmax \approx 0.00789$, $\rmax \approx 0.119$, and~$\dtot \approx 0.00385$.

As implied above, the presented approximations and bounds will be global ones, i.e., tight over the whole non-negative real axis (for all $x \geq 0$). The error functions converge to explicit values, which may be local extrema, at both ends of this range:
\begin{align}
\label{eq:limits}
 \lim_{\mathclap{x \to 0}} d(x) &=  ac - {\textstyle \frac{1}{2}},
&\:\:\:\:\lim_{\mathclap{x \to 0}} r(x) &= 2ac - 1,\\
\nonumber
 \lim_{\mathclap{x \to \infty}} d(x) &= 0,
&\:\:\:\:\lim_{\mathclap{x \to \infty}} r(x) &=
\begin{cases}
\infty,           & \text{if } b < {\textstyle \frac{1}{2}},\\
a\sqrt{2\pi} - 1, & \text{if } b = {\textstyle \frac{1}{2}},\\
-1,               & \text{if } b > {\textstyle \frac{1}{2}}.
\end{cases}
\end{align}
The last limit shows especially that global approximations and bounds in terms of relative error exist if and only if we set $b = \frac{1}{2}$. However, as a novel fact, our study demonstrates that absolute error and total error can be instead significantly reduced by permitting $b < \frac{1}{2}$. Therefore, two scenarios of approximations for the absolute and total error are considered in this Letter, i.e., approximations with $b=\frac{1}{2}$ or $b < \frac{1}{2}$.

Local error extrema may occur also at critical points, where the derivatives of the continuous error functions vanish. Denoting differentiation with an apostrophe, they are given by\\
$\displaystyle d'(x) = \tilde{Q}'(x) - Q'(x),\,\,\,
r'(x) = \frac{\tilde{Q}'(x)Q(x) -  \tilde{Q}(x)Q'(x)}{[Q(x)]^2},$\\
where
\begin{align}
\tilde{Q}'(x) &= -\dfrac{a\left(\left(2bx^2+1\right)\left(\mathrm{e}^{cx}-1\right)-cx\right)\mathrm{e}^{-bx^2-cx}}{x^2},\\
Q'(x) &= -\frac{1}{\sqrt{2\pi}} \exp\left(-{\textstyle \frac{1}{2}} x^2\right).
\end{align}

Two variations of approximations are considered herein: $d(0) = r(0) = 0$ and $d(0) = -\dmax$ (resp.\ $r(0) = -\rmax$). The former case maintains the continuity of the $Q$-function when extending to $x < 0$ and results in $c=\frac{1}{2\,a}$, when substituted in $\lim_{{x \to 0}} d(x)$ (resp.\ $\lim_{{x \to 0}} r(x)$) that is given in (\ref{eq:limits}). The latter case provides slightly better accuracy at the cost of discontinuity occurring at $x = 0$ and results in $c=\sqrt{\frac{\pi}{2}}$ in the cases of relative error, by solving $\lim_{{x \to 0}} r(x)=-\rmax$ with $\lim_{{x \to \infty}} r(x)=-\rmax$ that are defined in (\ref{eq:limits}). 

\section{Alternative Improved Coefficients for (\ref{eq:Q_and_Qtilde})}
\label{sec:theory}

In this section, we describe the methodologies to solve the new coefficients $(a,b,c)$ for the KL expression. They are optimized either in the minimax sense or in terms of the integrated total error to yield an approximation, an upper bound or a lower bound. All the 17 thus-obtained improved/alternative coefficient sets and accuracy thereof are listed in Table~\ref{tab:coefficients}.

\subsection{Global Uniform Approximations and Bounds}

The minimax optimization problems are solved in terms of both absolute and relative errors defined in (\ref{eq:abs_error}) and (\ref{eq:rel_error}), respectively, by constructing a set of nonlinear equations. This set describes the resulting error function, which should be uniform with equal values for all the extrema points. Each extremum point yields two equations, where one expresses its value and the other sets the derivative of the error function to zero at that point. In addition, one equation (for $d(x)$) or two equations (for $r(x)$) is/are obtained from evaluating the limits at the two endpoints of the considered range, $[0,\infty]$, per (\ref{eq:limits}).

The resulting sets of equations, which have equal number of equations and unknowns, can be solved straightforwardly by any numerical tool for the considered variations to find the optimized sets of coefficients that satisfy $\min_{(a,b,c)}\dmax$ for the absolute error and $\min_{(a,b,c)}\rmax$ for the relative error. We used iteratively random initial guesses for the unknowns in this approach, namely $(a,b,c)$, $\dmax$ or $\rmax$, and the location of the extrema ($x_k$), until {\tt fsolve} in Matlab converged to the solution, which is confirmed by substitution. The formulations for the minimax approximations/bounds are described below.

\subsubsection{Approximations in Terms of Absolute Error}

The coefficients $(a,b,c)$ are optimized for approximations in terms of the absolute error by formulating a set of equations as 
\begin{align}
\label{equ:apr-abs}
	\begin{cases}
	    d'(x_k) = 0,&\hspace{-2.5cm}\text{for } k=1,2 \text{ or } 1,2,3,\\
	    d(x_k) = (-1)^{k+1}\,\dmax,&\hspace{-2.5cm}\text{for } k=1,2 \text{ or } 1,2,3,\\
	    \begin{cases}
	        a\,c=\frac{1}{2},&\text{when }d(0)=0,\\
	        a\,c=\frac{1}{2}-\dmax,&\text{when }d(0)=-\dmax,
	    \end{cases}
	\end{cases}
\end{align}
where $x_k$ is an extremum point. The number of the error function's extrema depends on the value of $b$; if $b$ is fixed to $\frac{1}{2}$, then we have two extrema, whereas if $b$ is allowed to be any positive value, then we need three separate extrema.

\subsubsection{Lower Bounds in Terms of Absolute Error}

For the lower bounds, we need to find the optimized coefficients which minimize the global absolute error for $d(x)\leq 0$ when $x\geq 0$. The value of $b$ must always equal to $\frac{1}{2}$. The tightest resulting uniform error function will start from $d(0)=-\dmax$, with its maximum equal to zero and its minimum equal to $-\dmax$ so that we can formulate a set of equations as 
\begin{align}
\label{abs_lower}
	\begin{cases}
	    d'(x_1) = d'(x_2) = 0,\\
	    d(x_1) = 0,\,
	    d(x_2) =-\dmax,\\
	    a\,c=\frac{1}{2}-\dmax.
	\end{cases}
\end{align}
When $d(0)=0$, we get $a=\sqrt{\frac{\pi}{32}}$ and $c=\sqrt{\frac{8}{\pi}}$ by imposing $d'(0)=0$ (only in this case), which produces $a\,c^2=\sqrt{2/\pi}$, and solving with $c=\frac{1}{2\,a}$ that results from setting $d(0)=0$.

\subsubsection{Upper Bounds in Terms of Absolute Error}

The set of equations becomes
\begin{align}
\label{upper_abs}
	\begin{cases}
	    d'(x_1) = d'(x_2) = d'(x_3) = 0,\\
	    d(x_1) = d(x_3) = \dmax,\,
	    d(x_2) = 0,\\
	    a\,c=\frac{1}{2}.
	\end{cases}
\end{align}
In particular, we shape the uniform error function to have three extrema with $d(x)\geq 0$ when $x\geq 0$ in which its maxima are equal to $\dmax$ and its minimum is equal to zero. The corresponding error function must always start from $d(0)=0$.

\subsubsection{Approximations in Terms of Relative Error}

The targeted uniform error function in terms of the relative error consists of only one maximum point and converges to $-\rmax$ as $x$ tends to infinity, which results in $-\rmax=a\sqrt{2\,\pi}-1$ according to (\ref{eq:limits}). Therefore, we can formulate the set of equations as 
\begin{align}
\label{equ:abr_rel}
	\begin{cases}
	    r'(x_1) = 0,\,
	    r(x_1) = \rmax,\\
	    \begin{cases}
	        a\,c=\frac{1}{2},&\text{when }r(0)=0,\\
	        a\,c=\frac{1-\rmax}{2},&\text{when }r(0)=-\rmax,
	    \end{cases}\\
	    a=\frac{1-\rmax}{\sqrt{2\,\pi}}.
	\end{cases}
\end{align}

\subsubsection{Lower Bounds in Terms of Relative Error}

We need to find the optimized coefficients, $a$ and $c$, in the minimax sense for $r(x)\leq 0$ when $x\geq 0$ which converges to $-\rmax$ as $x$ tends to infinity. The resulting error function can either start from $r(0)=-\rmax$ to formulate a set of equations as
\begin{align}
\label{rel_lower}
	\begin{cases}
	    r'(x_1) = r(x_1) = 0,\\
		a\,c=\frac{1-\rmax}{2},\,
		a=\frac{1-\rmax}{\sqrt{2\,\pi}},
	\end{cases}
\end{align}
or from $r(0)=0$ yielding $a=\sqrt{\frac{\pi}{32}}$ and $c=\sqrt{\frac{8}{\pi}}$ like with the corresponding lower bound in terms of absolute error.

\subsubsection{Upper Bound in Terms of Relative Error}

We must ensure that $r(x)\geq 0$ when $x\geq 0$ for the uniform error function. The resulting error function has only one maximum point and converges to zero as $x$ tends to infinity. Therefore, $a=\frac{1}{\sqrt{2\pi}}$ and $c=\sqrt\frac{\pi}{2}$ as proposed earlier in~\cite{Jang-2011Feb} and $b$ is known to be equal to $\frac{1}{2}$. The optimized upper bound in terms of relative error is also optimal in terms of absolute error and integrated total error for the case where $b=\frac{1}{2}$.

\subsection{Numerical Optimization in Terms of Total Error}

Instead of defining $\dtot \triangleq \int_{0}^{R}|d(x)|\,\mathrm{d}x$ like in \cite{Karagiannidis-Lioumpas-2007Aug} and so making optimized coefficients specific to the value chosen for $R$ and limited to the range $[0,R]$, we measure total error with $R\to\infty$ and obtain globally optimized approximations and bounds. In particular, we optimized the coefficients for the two variations of the approximations with or without setting $b=\frac{1}{2}$ by performing an extensive search, where we evaluated the target metric ($\dtot$) over wide one/two/three-dimensional grids for the unknowns $a$, $(a,b)$, $(a,c)$, or $(a,b,c)$ with granularity of $0.000001$ and selected the grid point with the minimum total error for each variation. This renders four sets of optimized coefficients. Extra constraint checks guarantee $d(x)<0$ for the lower bound and $d(x)>0$ for the upper bound.

\section{Numerical Results and Conclusions}
\label{sec:NumericalResults}

\begin{table}
    \begin{center}
    \caption{New coefficients for (\ref{eq:Q_and_Qtilde}) and approximation error thereof}\label{tab:coefficients}
    \includegraphics[scale=1.00]{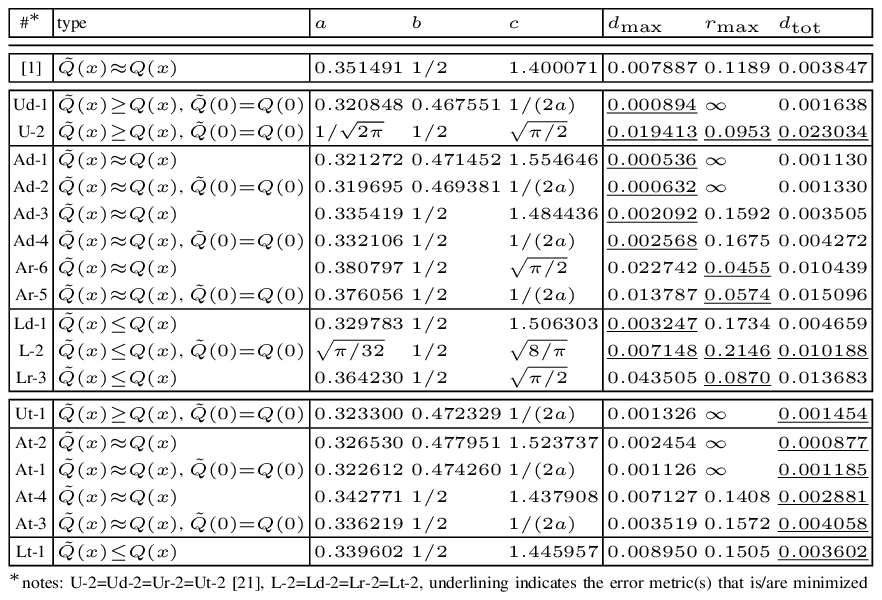}
    \end{center}
\end{table}

\begin{figure*}
	\begin{center}
	    \includegraphics[scale=1.00]{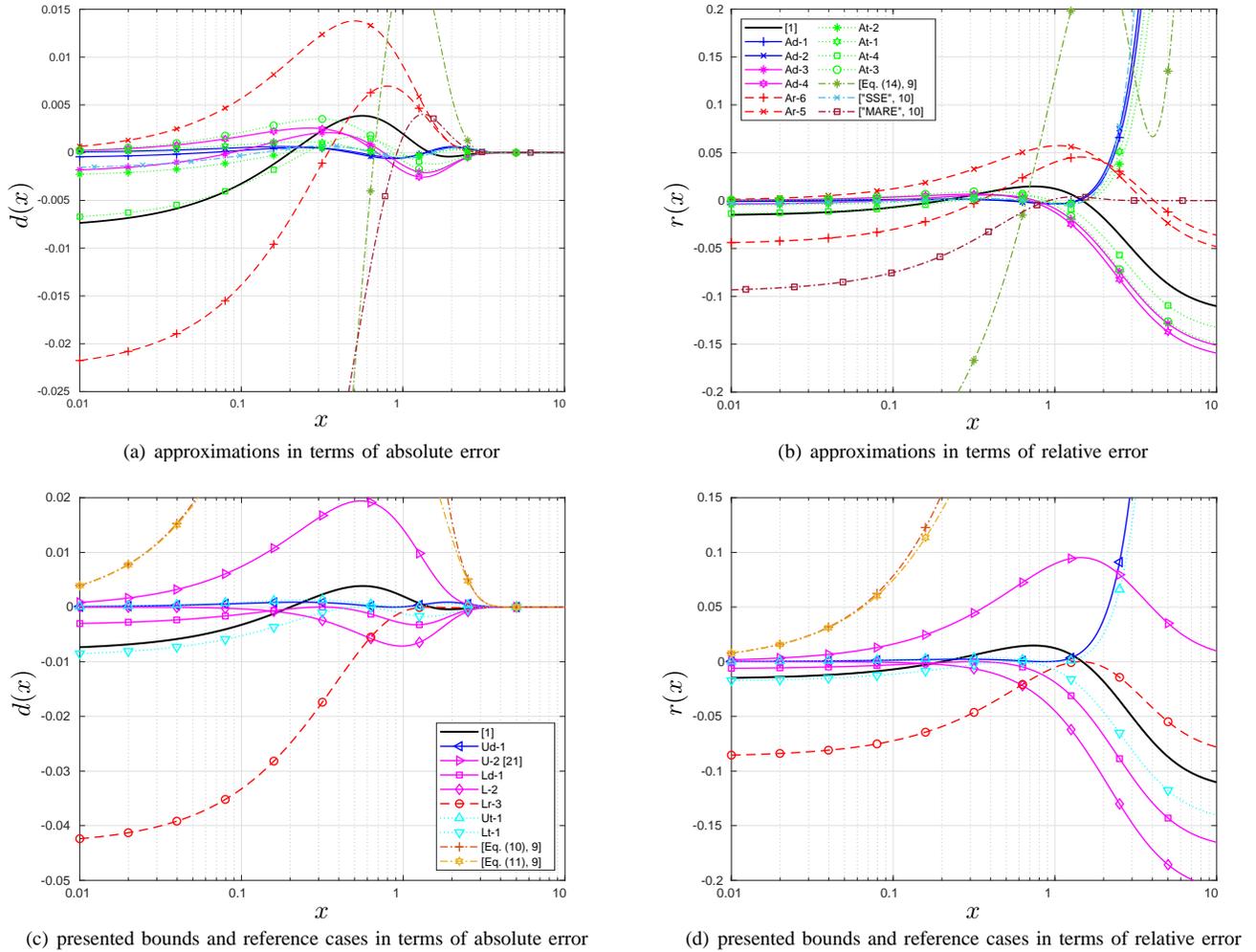}
		\caption{The improved approximations and bounds compared to the KL approximation with the original coefficients~\cite{Karagiannidis-Lioumpas-2007Aug} and to expressions from \cite{Chiani-Dardari-Simon-2003Jul} and \cite{LopezBenitez-Casadevall-2011Apr}.}\label{fig:fig2}
    \end{center}
\end{figure*}

We summarize the improved coefficients for the minimax approximations and bounds and for the total absolute error in Table~\ref{tab:coefficients} and illustrate their error functions in Fig.~\ref{fig:fig2}, together with the original KL approximation from\cite{Karagiannidis-Lioumpas-2007Aug} and reference approximations and bounds from \cite{Chiani-Dardari-Simon-2003Jul} and \cite{LopezBenitez-Casadevall-2011Apr}.\footnote{The labels having the form $Xy$-$n$ in the results refer to the approximations and bounds as follows: $X$ is U for upper bounds, A for approximations, and L for lower bounds; whereas $y$ is d for absolute error, r for relative error, and t for total error; in addition, $n$ refers to rank of the coefficients according to the accuracy of the absolute error of each variation in an ascending order.}

The numerical results show that the improved coefficients of the proposed KL approximations and bounds are optimal subject to their optimization targets, yet expressed precisely in implicit form as solutions to systems of nonlinear equations as opposed to relying on numerical search to minimize error measures. In some specific cases, a part or even all of the three coefficients can be expressed as explicit constants. The best approximation/bound from Table~\ref{tab:coefficients} for a specific application is chosen by contrasting requirements against Fig.~\ref{fig:fig2}, provided that the KL expression (\ref{eq:Q_and_Qtilde}) is suitable for it to begin with.

As an ultimate conclusion, the presented data suggests good alternatives to the original coefficients given in~\cite{Karagiannidis-Lioumpas-2007Aug} for the case of $b=\frac{1}{2}$: In some applications, the accuracy of the KL approximation might be improved by choosing instead $A = 1.95$, $B = 1.113$ (a compromise between all \mbox{A$y$-$n$}) for decreasing both absolute and relative error by round $15$\% at the cost of increasing total error by round $65$\%; or $A = 2.03$, $B = 1.162$ (At-4) for decreasing absolute error and total error by round $10$\% and $25$\%, respectively, at the cost of increasing relative error by round $15$\%. Sometimes it may also be useful to choose $A = B\sqrt{\pi} \approx 1.88$, $B = 1.061$ (Ar-5) for minimizing relative error (with round $50$\% reduction) subject to zero error at the origin. In contrast, when primarily minimizing absolute error, accuracy can be improved significantly by generalizing the KL approximation to allow any positive $b$: Namely, the choice $a = 0.32$, $b = 0.4703$, $c = 1.5625$ (Ad-2) guarantees zero error at the origin while decreasing absolute error and total error as much as round $90$\% and $65$\%, respectively, at the cost of making relative error unbounded for large arguments.

\IEEEtriggeratref{13}

\end{document}